\documentstyle[12pt]{article}

\def\reference{\parskip 0pt\par\noindent\hangindent 0.5 truecm}

\begin{document}

\title{Gravitational lensing and modified Newtonian dynamics}

\author{
	Daniel J.\ Mortlock$^{1,2}$ \and
 	Edwin L.\ Turner$^{3}$
}

\date{}
\maketitle

{\center
	$^1$ Astrophysics Group, Cavendish Laboratory, Madingley Road,
	Cambridge CB3 0HE,
        United Kingdom \\
	mortlock@ast.cam.ac.uk\\[3mm]
	$^2$ Institute of Astronomy, Madingley Road, Cambridge CB3 0HA,
	United Kingdom \\[3mm]
	$^3$ Princeton University Observatory, Peyton Hall,
	Princeton, NJ 08544, U.S.A.\ \\
	elt@astro.princeton.edu\\
}

\begin{abstract}
Gravitational lensing is most often used as a tool to
investigate the distribution of (dark) matter in the universe, 
but, if the mass distribution is known a priori,
it becomes, at least in principle, a powerful probe of gravity itself.
Lensing observations are a more powerful tool than dynamical measurements
because they allow measurements of the gravitational field far
away from visible matter.
For example, modified Newtonian dynamics (MOND) has no relativistic
extension, and so makes no firm lensing predictions, but
galaxy-galaxy lensing data can be used to empirically constrain the
deflection law of a MONDian point-mass.
The implied 
MONDian lensing formalism is consistent with general relativity, in so far
as the deflection experienced by a photon is twice that experienced
by a massive particle moving at the speed of light.
With the deflection law in place and no invisible matter,
MOND can be tested wherever lensing is observed.
\end{abstract}


\bigskip

\section{Introduction}
\label{section:intro}

In the time since the discovery of the first multiply-imaged
quasar (Walsh, Carswell \& Weymann 1979), 
gravitational lensing has become one of the more powerful tools
available to astronomers.
The main advantage of lensing as an astronomical probe lies in 
its simplicity -- the light deflection properties of a lens
depend on just two things: its mass distribution and the 
nature of gravitational physics.

The success of general relativity (GR)
-- and, in the appropriate limits, Newtonian gravity --
is such that lensing is almost always used to
investigate the distribution of matter in the universe.
The fact that, for example, cluster lensing cannot be explained
by the visible galaxies and gas is then taken to be strong
evidence for the preponderance of dark matter.
There is, of course, a great deal of evidence that the universe 
is dark matter-dominated (e.g., Trimble 1987), but the lack of 
any non-gravitational detection of dark mass makes it an assumption
that should continue to be tested by all available means.

The alternative possibility is that GR is only
valid in the small-scale, large-acceleration regimes in which it
has been experimentally (as opposed to observationally) tested.
The majority of direct tests have been within the Solar system, but 
measurements of time delays in binary pulsar systems 
(Taylor et al.\ 1992) 
have also verified GR. 
The degeneracy between the form of the gravitational acceleration
and the distribution of mass is such 
that GR remains an assumption on galactic (and greater) scales.
If it is hypothesised that there is no
dark matter, then it is possible to determine the nature of 
gravity on these large scales from lensing or dynamical observations.
Aside from their tendency to rely on assumptions of equilibrium,
dynamical measurements are subject to the more fundamental limitation
that the gravitational field can only be probed in regions where there
is visible matter. By contrast lensing can be
used to measure gravitational effects well beyond the luminous
extent of the deflector(s). 
Further, if such measurements can be made sufficiently far from the
lens\footnote{Note that, in the absence of dark matter, it is only 
the visible extent of the deflector that is relevant.}, 
its internal structure becomes unimportant, and the lensing 
data uniquely constrains the deflection law of a point-mass.
This is a potentially powerful technique that only breaks down
at angular scales so large that the effects of other deflectors
along the line-of-sight become important.

These methods can be used to
determine the deflection law of a point-mass empirically
(Mortlock \& Turner 2001a) or to test a particular
theory. For example Mortlock \& Turner (2001b)
investigated gravitational lensing within the
framework of modified Newtonian dynamics (MOND).
A combined approach is taken here, using MOND as an illustrative
example.
After explaining the principles of the theory (Section~\ref{section:mond}),
a robust deflection law is derived from galaxy-galaxy lensing data
(Section~\ref{section:obs}).
The future prospects for more rigorous tests are then discussed
in Section~\ref{section:conc}.

\section{MOND}
\label{section:mond}

MOND (Milgrom 1983) hypothesises that the inertial mass of 
a particle is decreased when it is subject to an acceleration much 
weaker than a critical value, 
$a_0 \simeq 1.2 \pm 0.1 \times 10^{-10}$ m s$^{-2}$. 
In its purest form a MONDian universe contains no dark matter,
and, remarkably, this simple model can explain
the dynamics of galaxies and clusters (McGaugh \& de Blok 1998),
as well as the observed power spectrum of cosmic microwave background
anisotropies (McGaugh 2000). 

MOND is, of course, highly unconventional and 
the variation of inertia would be very difficult to 
integrate into the current broader understanding of the physical world.
This fact alone is sufficient to convince many that it is extremely unlikely
to be correct.  
However, even if one takes this (anti-empirical) point of
view, MOND provides a potentially revealing opportunity to test the depth
and robustness of modern science's observational knowledge of the universe.  
If it is not possible to contradict such a simple but seemingly
outlandish and improbable hypothesis, 
how much confidence should be placed in more conventional
explanations of the observed universe?

Another difficulty with MOND is that it is not a complete 
physical theory, lacking a relativistic extension,
and thus making no definite predictions for light deflection
(Milgrom 1983; Bekenstein \& Milgrom 1984).
The natural Ansatz for MONDian lensing is, as in GR,
that a photon experiences twice the deflection of a massive particle
moving at the speed of light (Qin, Wu \& Zou 1995). 
This hypothesis gives qualitatively reasonable predictions 
(Mortlock \& Turner 2001), but the effects of 
extended and multiple deflectors are somewhat ambiguous.
However, the gravitational properties of an isolated point-mass, $M$,
are well defined: the effective force law
matches the Newtonian form for $r \ll r_{\rm M} =
(G M / a_0)^{1/2}$, but falls off as $r^{-1}$ for $r \gg r_{\rm M}$.
The details of the physics for $a \simeq a_0$ are unspecified, but
unimportant in the absence of high precision measurements.
Integrating this acceleration along the line-of-sight gives
the (reduced)
bending angle, $\alpha(\theta)$, which matches the Schwarzschild 
form for small impact parameters [i.e., $\alpha(\theta) \propto \theta^{-1}$],
but is constant beyond $\theta = r_{\rm M} / d_{\rm od}$.
(Here $d_{\rm od}$ is the angular diameter distance from observer
to deflector, which is not formally defined in MOND.) 
The deflection angle is not directly measurable, but the distortion
of images and the (total)
magnification of sources can be calculated directly from 
$\alpha(\theta)$, and these are observable.

\section{Observational constraints}
\label{section:obs}

Gravitational lensing observations range from light deflection
by the Sun, through microlensing in the local group, to the multiple
imaging of high-redshift sources. If it is assumed that there is 
no dark matter then all of these observations place constraints on 
the nature of gravitational light deflection, but only some represent clean
and powerful probes of MOND, whereas others are clearly within the Newtonian
regime, or require the untangling of the combined effects of multiple
deflectors. These issues are explored more fully by 
Mortlock \& Turner (2001a), but it is clear that
galaxy-galaxy lensing observations offer by far the best 
opportunity to make interesting inferences from available data.

Galaxy-galaxy lensing, the weak tangential alignment of distant 
galaxies caused by their more nearby counterparts, has been used to 
weigh the dark matter distributions of the foreground deflectors
(e.g., Brainerd, Blandford \& Smail 1996; Fischer et al.\ 2000).
The results are all consistent with galaxies having
large isothermal haloes
extending to at least several hundred kpc.
Interestingly, no upper limit has been placed on the halo size,
despite the fact that a systematic distortion has been
measured out to $\sim 1000$ arcsec,
far beyond the visible extent of the foreground galaxies (only a 
few arcsec).
Thus, under the no-dark matter hypothesis, 
these data represent an ideal 
means of measuring the deflection law of what is effectively
an isolated point-mass.
Even without performing any further analysis, the fact that the
lensing data are consistent with rotation curve measurements 
in GR implies that the relationship between the deflection of 
massive and massless particles must be the same in MOND (or any
other theory). 

\begin{figure}
\includegraphics{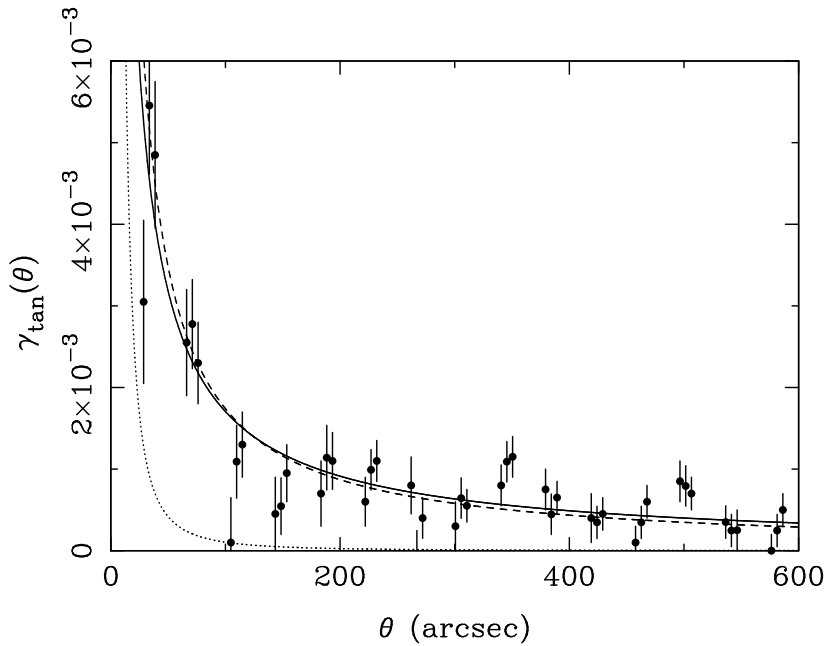}
\vspace{93mm}
\caption{The mean shear, $\gamma_{\rm tan}(\theta)$,
around foreground galaxies
in the $g^\prime$, $r^\prime$ and $i^\prime$ bands,
as measured by Fischer et al.\ (2000)
is compared to various theoretical
predictions.
In each bin the data in the three bands (which are offset for clarity)
are strongly correlated as the errors are dominated by the sample 
noise in the orientations of the background galaxies.
The three models shown are:
the best-fit power law (solid line);
the MONDian prediction (dashed line);
and the Newtonian result if there is no dark matter (dotted line).}
\label{figure:galgal}
\end{figure}

More quantitatively, Fischer et al.\ (2000) fit 
the shear signal around $\sim 3 \times 10^4$ 
foreground galaxies by 
\begin{equation}
\gamma_{\rm tan}(\theta) = \gamma_{\rm tan,60} 
\left(\frac{60 \,\, {\rm arcsec}}{\theta}\right)^\eta,
\end{equation}
with $\gamma_{\rm tan,60} = 0.0027 \pm 0.0005$ 
and 
$\eta = 0.9 \pm 0.1$.
Figure~\ref{figure:galgal} shows this fit to the data,
along with 
the predictions of GR (assuming no dark matter)
and the MONDian lensing formalism described in Section~\ref{section:mond}.
The MONDian results (which imply $\eta = 1$) are clearly consistent with the 
data,\footnote{This agreement also provides quantitative support for
the hypothesis that, in MOND (or indeed any alternative gravity 
theory), photons experience twice the deflection of massive
particles moving at the speed of light.}
although there is some ambiguity in the normalisation,
as the mass-to-light ratio
of the deflectors are not known with great certainty 
(Mortlock \& Turner 2001b).
Thus the properties of MONDian point-mass lenses must be
considered completely (if approximately) defined.
It is not clear, however, that the entire derivation of
the deflection law given in Mortlock \& Turner (2001b) is
verified, and so the lensing properties of more complex
deflectors remain unknown, although future investigations
should shed some light on this matter as well.

\section{Conclusions}
\label{section:conc}

MOND, an alternative theory of gravity (or inertia), 
has been able to explain the dynamics of massive 
bodies in terms of visible matter where Newtonian 
physics requires large amounts of dark matter. 
However MOND has no relativistic extension and so makes
no predictions about gravitational lensing. 
Here the opposite approach has been taken, using 
observations of lensing to constrain the form of 
MONDian light deflection. 
The cleanest way of doing this is to use galaxy-galaxy
lensing data to show that the deflection of photons is 
simply twice the deflection that would be experienced by a
massive particle moving at the speed of light.
Thus, a relativistic MONDian theory must have a great deal in common
with GR.

With one more part of a complete theory in place a number of 
new observational tests become available. 
The clearest relate to ``simple'' lenses: situations in which there
is a single, isolated deflector that has no internal structure on the
scales of interest. 
Microlensing within the local group fulfills these criteria, although
the typical impact parameters are so small that it is
only the low-magnification tails of the light-curves that are of 
interest. However microlensing by cosmologically distant deflectors
offers a good opportunity to confirm or reject MOND. 
Several programmes to measure low optical depth microlensing of 
high-redshift quasars are underway (Walker 1999; Tadros, Warren \&
Hewett 2001),
but thus far no events have been observed. 
The only (probable) example of cosmological microlensing
by an isolated deflector observed to
date is the serendipitous detection of a peak
in the light curve of gamma ray burst 000301C 
(Garnavich, Loeb \& Stanek 2000), but the photometry 
is not of sufficient quality to differentiate between MOND and GR.

Further observations of
galaxy-galaxy lensing
offer the most
likely tests of MONDian lensing in the near future.
Any asymmetry in the distortions (c.f.\ Natarajan \& Refregier 2000)
would be at odds with a no-dark matter theory,
and there is also the possibility of measuring an outer cut-off
in the shear signal. This must eventually come from the influence of
secondary deflectors along the line-of-sight, but could also signify
a putative return to a Newtonian regime in the ultra-weak acceleration
limit.
If MOND is not contradicted by any of the above observations, more
complex situations should be investigated. Multiple or extended
deflectors must also be able to explain observed shear fields and cases
of multiple imaging without recourse to invisible mass.

\section*{Acknowledgements}

DJM is funded by PPARC, and
this work was supported in part by NSF grant AST98-02802.

\section*{References}

\reference Beckenstein, J., Sanders, R.H., 1994, ApJ, 429, 480
\reference Brainerd, T.G., Blandford, R.D., Smail, I.S., 1996, ApJ, 466, 623
\reference Fischer, P., et al., 2000, AJ, 120, 1198
\reference Garnavich, P.M., Loeb, A., Stanek, K.Z., 2000, ApJ, 544, L11
\reference McGaugh, S.S., 2000, ApJ, 541, L33
\reference McGaugh, S.S., de Blok, W.J.G., 1998, ApJ, 499, 66
\reference Milgrom, M., 1983, ApJ, 270, 365
\reference Mortlock, D.J., Turner, E.L., 2001a, MNRAS, submitted
\reference Mortlock, D.J., Turner, E.L., 2001b, MNRAS, submitted
\reference Natarajan, P., Refregier, A., 2000, ApJ, 538, L113
\reference Trimble, V., 1987, ARA\&A, 25, 425
\reference Qin, B., Wu, X.P., Zou, Z.L., 1995, A\&A, 296, 264
\reference Sanders, R.H., 1994, A\&A, 284, L31
\reference Walker, M.A., 1999, MNRAS, 306, 504
\reference Tadros, H., Warren, S.J., Hewett, P.C., 2001, in
        Cosmological Physics with Gravitational Lensing,
        eds.\ Kneib, J.-P., Mellier, Y., Moniez, M., Tran Thanh Van, J.,
        Edition Frontiers, in press
\reference Taylor, J.H., Wolszczan, A., Damour, T., Weisberg, J.M.,
       1992, Nature, 355, 132
\reference Walsh, D., Carswell, R.F., Weymann, R.J., 1979, Nature, 279, 381

\end{document}